\documentclass[traditabstract]{aa}  
\usepackage{graphicx}
\usepackage{epstopdf}
\usepackage{natbib}
\usepackage{txfonts}
\usepackage{color}

\begin{document}
   \titlerunning{Radio Observations of Bi-directional Electron Beams}
   \authorrunning{E. Carley et al.}
   \title{Low frequency radio observations of bi-directional electron beams in the solar corona}

   \author{Eoin P. Carley
          \inst{1,2},
          Hamish Reid
          \inst{3},
          Nicole Vilmer
          \inst{2}
          \and
          Peter T. Gallagher\inst{1}
          }

   \institute{Astrophysics Research Group, School of Physics, Trinity College Dublin, Dublin 2, Ireland.\\
 			\and
			LESIA, Observatoire de Paris, CNRS, Univ. Paris 6 \& 7, 92195 Meudon, France. \\
   			\and
   			SUPA School of Physics \& Astronomy, University of Glasgow, G12 8QQ, United Kingdom.\\
		   \email{eoin.carley@obspm.fr}
                    }

   \date{Submitted 2 April 2015; Accepted 2 August 2015}

 
  \abstract{
The radio signature of a shock travelling through the solar corona is known as a type II solar radio burst. In rare cases these bursts can exhibit a fine structure known as `herringbones', which are a direct indicator of particle acceleration occurring at the shock front.
%
However, few studies have been performed on herringbones and the details of the underlying particle acceleration processes are unknown.
Here, we use an image processing technique known as the Hough transform to statistically analyse the herringbone fine structure in a radio burst at $\sim$20--90\,MHz observed from the Rosse Solar--Terrestrial Observatory on 2011 September 22.
We identify 188 individual bursts which are signatures of bi-directional electron beams continuously accelerated to speeds of 0.16$_{-0.10}^{+0.11}\,c$. This occurs at a shock acceleration site initially at a constant altitude of $\sim$0.6 R$_{\odot}$ in the corona, {\color{black} followed by a shift to $\sim$0.5 R$_{\odot}$}. The anti-sunward beams travel a distance of 170$_{-97}^{+174}$\,Mm {\color{black}(and possibly further)} away from the acceleration site, while those travelling toward the sun come to a stop sooner, reaching a smaller distance of 112$_{-76}^{+84}$\,Mm.
We show that the stopping distance for the sunward beams may depend on the total number density and the velocity of the beam.
Our study concludes that a detailed statistical analysis of herringbone fine structure can provide information on the physical properties of the corona which lead to these {\color{black}relatively} rare radio bursts.}


   \maketitle
%


\section{Introduction}
The longest known signature of shocks propagating into the solar corona are type II radio bursts \citep{wild1958, nelson1985}. It has been postulated that they are produced by a disturbance travelling into the solar atmosphere at speeds of over 500\,km\,s$^{-1}$. About 20\% of type IIs contain a fine structure of pulsations or `thorns', known as herringbones \citep{roberts1959}, {\color{black}which typically occur between 10--120\,MHz \citep{cairns1987}}, and are thought to be observations of electron beams escaping the shock as it propagates. Although herringbone radio bursts provide a wealth of information on the particle acceleration process inside coronal shocks, few studies of their properties have been made. A statistical analysis of herringbone features could yield important information on the type of acceleration process occurring in coronal shocks and/or the environment into which particles are accelerated. 

Observations of herringbone bursts were first reported by \citet{roberts1959} and a description was later provided by \citet{wild1958} whereby \emph{`shock fronts propagate normal to the magnetic field and continuously eject bunches of fast electrons along the field'}; this assertion was also illustrated in \citet{stewart1980}. {\color{black} Although herringbones appear quite similar to bi-directional type IIIs, such as those reported in \citet{aschwanden1995}}, \citet{cairns1987} concluded that herringbones are a different phenomenon to type III bursts; {\color{black} their frequency-time profiles are characteristically different, $\sim$80\% of herringbones have stronger fundamental emission than harmonic (it is generally the opposite in type IIIs), and
the sense of polarisation of their fundamental and harmonic bands is opposite to that of type IIIs. 
Indeed, later findings eventually confirmed a closer link to shock activity in the corona, with more intense type IIs being more likely to have herringbone structure \citep{cane1989}.} Regarding the properties of shock accelerated particles, \citet{mann1995} and \citet{mann2005} showed that the electron beams that cause herringbones have speeds of $\sim$0.1\,c. 
\citet{carley2013} reported similar speeds and showed that the acceleration process has a quasi-periodicity of between 2-11 seconds and may be due to particle acceleration at an expanding CME flank. 

There has been little theoretical investigation as to the origin of the bursts. For example, what gives herringbone radio bursts this quasi-periodicity or `burstiness' in time? And what are the coronal structures that lead to herringbones? \citet{zlobec1993} and later \citet{vandas2011} suggested that particles accelerated in a magnetic trap on a wavy shock front might be responsible. Similarly, it has been suggested that the burstiness is due to inhomogeneity and repeated electron acceleration on a wavy shock \citep{guo2010, burgess2006}. \citet{schmidt2012} presented such a scenario in a model of a rippled shock on a CME flank accelerating particle beams into the surrounding corona. {\color{black} Alternatively, herringbones may be produced in the termination shocks of super-Alfv\'{e}nic outflow jets of a reconnection region \citep{aurass2002, mann2009}. Such a theory may explain why some herringbones appear to come from a stationary height in the corona \citep{aurass2004}.}

Despite the combination of observation, theory, and modelling that has gone before, there is no complete explanation of herringbone fine structure. An understanding of these bursts could benefit from a statistical analysis such as was previously performed for type IIIs {\color{black}\citep{Saint-Hilaire2013, lobzin2011}}. However, because of the high number of individual herringbones in a typical radio burst, an automatic feature recognition technique may be required to study them accurately. One such technique is the Hough transform \citep{hough1961, duda1972}, used to identify straight lines in images. This tool was employed as a feature recognition technique for type IIIs as part of the HELIO Feature Catalogue \citep{bentley2011, bonnin2011} and also for automatic recognition of type IIs \citep{lobzin2010}. Those studies were for large data sets; here we constrain ourselves to a single radio event exhibiting herringbone bursts to test the applicability of the Hough transform to this kind of radio fine structure. 

Section 2 concentrates on radio observations of the event; Section 3 provides a description of the Hough transform and methods; Section 4 gives a statistical analysis of herringbone burst properties; and Sections 5 and 6 provide a discussion of the physical interpretation of the bursts and conclusions, respectively.


\section{Observations}

\begin{figure}[!h]
    \begin{center}
    \includegraphics[scale=0.38, trim = 1.1cm 0cm 0cm 0cm]{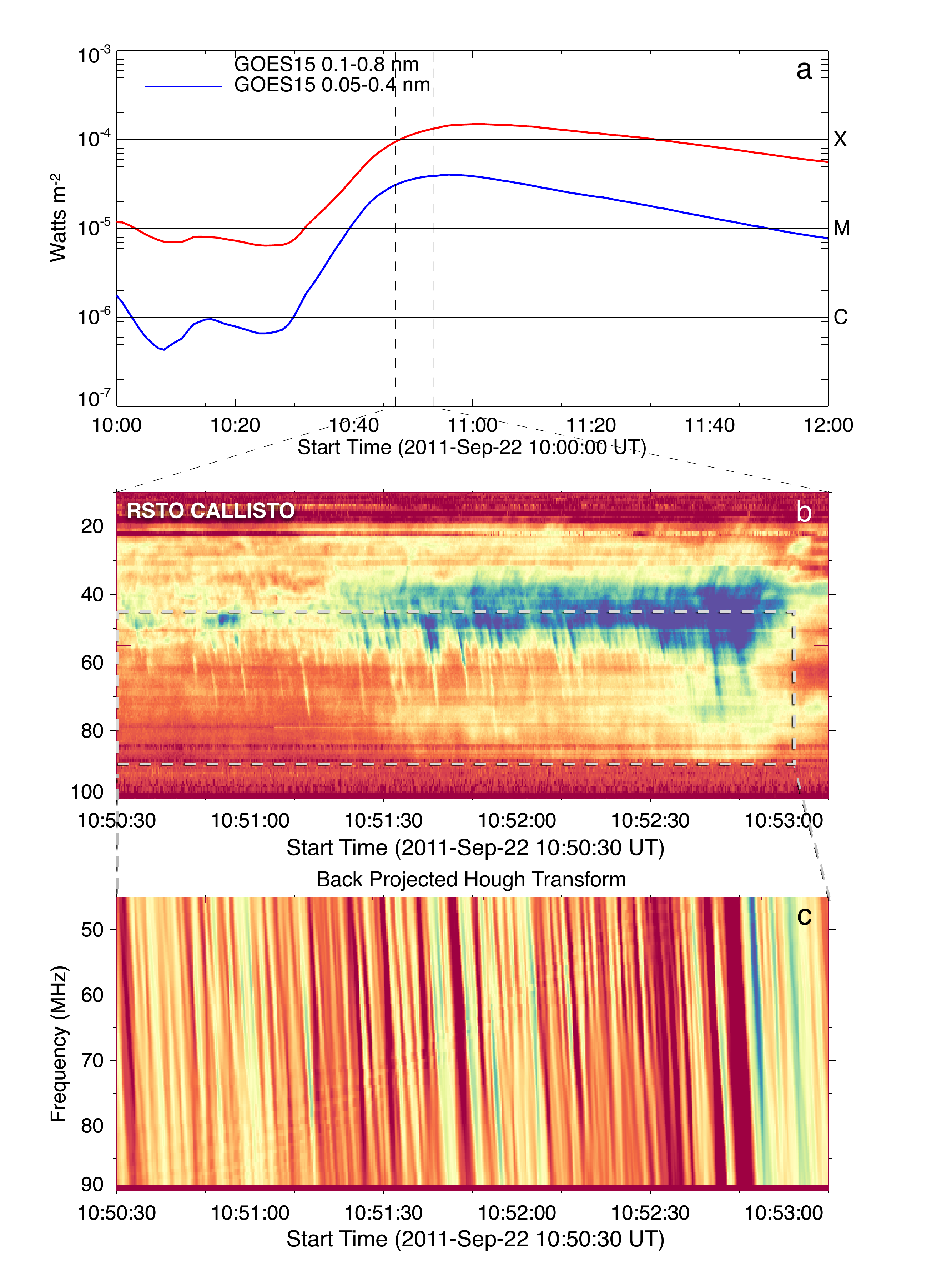}
    \caption[Callisto observations type II and herringbones]{ {\color{black}(a) GOES soft X-ray light curves showing an X1.4 class flare on 22 September 2011 at $\sim$10:30\,UT. (b) During the impulsive/peak phase of the flare the RSTO 10--100\,MHz Callisto receiver observed herringbone radio bursts from $\sim$10:48--10:55\,UT; only a section of all herringbone activity is shown here for clarity. (c) The back projected Hough transform of the area containing the reverse drift bursts. This data contains positions of all bursts, which are much more clearly defined and subject to less RFI or background noise. {\color{black} We note that the Hough transform provides the positions of burst maxima but does not produce the stop frequency in the dynamic spectra. For our analysis the stopping frequency was chosen manually. } For a complete overview of all radio activity from this event, see \citet{carley2013}}.}
    \label{fig:goes_hb}
    \end{center}
\end{figure}

{The SOL2011-09-22T10:30 event (X1.4, Figure~\ref{fig:goes_hb}a)} was accompanied by a variety of radio activity including type IIs, type IIIs and herringbone radio bursts. The flare occurred at N09E89 and was accompanied by a CME, EUV wave and a number of radio sources imaged by the Nan\c{c}ay Radioheliograph {\color{black} \citep[NRH;][]{kerdraon1997}}; a detailed analysis of the event is given in \citet{carley2013}. The herringbone radio activity began at $\sim$10:47:30\,UT between 10-90\,MHz, detected using the lowest frequency Callisto receiver at the Rosse Solar--Terrestrial Observatory \citep[RSTO;][]{zucca2012}, at Birr Castle, Ireland. Herringbones drifting to both high (reverse drift) and low frequency (forward drift) can be seen in Figure~\ref{fig:goes_hb}b. An interesting feature of this burst was the lack of drift of the burst `backbone' i.e. the herringbones appear to originate from the same frequency over time, similar to observations reported in {\color{black}\citet{aurass2002, melnik2004, melnik2005}}. This is suggestive of particle acceleration from a shock at a constant altitude, or at least in an environment of constant density in the frame of the shock. Usually, the backbone drifts to low frequencies like a normal type II while emitting herringbones (see \citet{miteva2007} for a good example). 

A previous analysis showed that there is some level of periodicity of between 2--11 seconds on the occurrence of the bursts in this event \citep{carley2013}. Our goal here is to further statistically analyse the bursts in terms of drift rate and intensity as well as the velocity and {\color{black} distance travelled} of the electron beams causing the radio emission.

 
\section{Method}

To perform a statistical analysis we used the Hough transform \citep{hough1961}, a feature recognition algorithm that is used to identify straight lines in images. In the Hough transform any point in an image may be represented in a polar coordinate `Hough space' as a sinusoid; if two points in the original image lie on a straight line, their sinusoids intersect in the Hough space. Lines in the original image space are found by searching for the intersection points of sinusoids in Hough space, see \citet{duda1972} for more details. If the image is transformed into Hough space, it is possible to `back-project' the Hough space to re-obtain the original image. Furthermore, it is possible to back-project a particular region of the Hough space, such that only lines with a particular orientation in the original image will pass through the transform. This is very useful for filtering an image for linear features of a particular orientation or slope.

\begin{figure}[!t]
    \begin{center}
    \includegraphics[scale=0.35, trim=1cm 11cm 0cm 0cm]{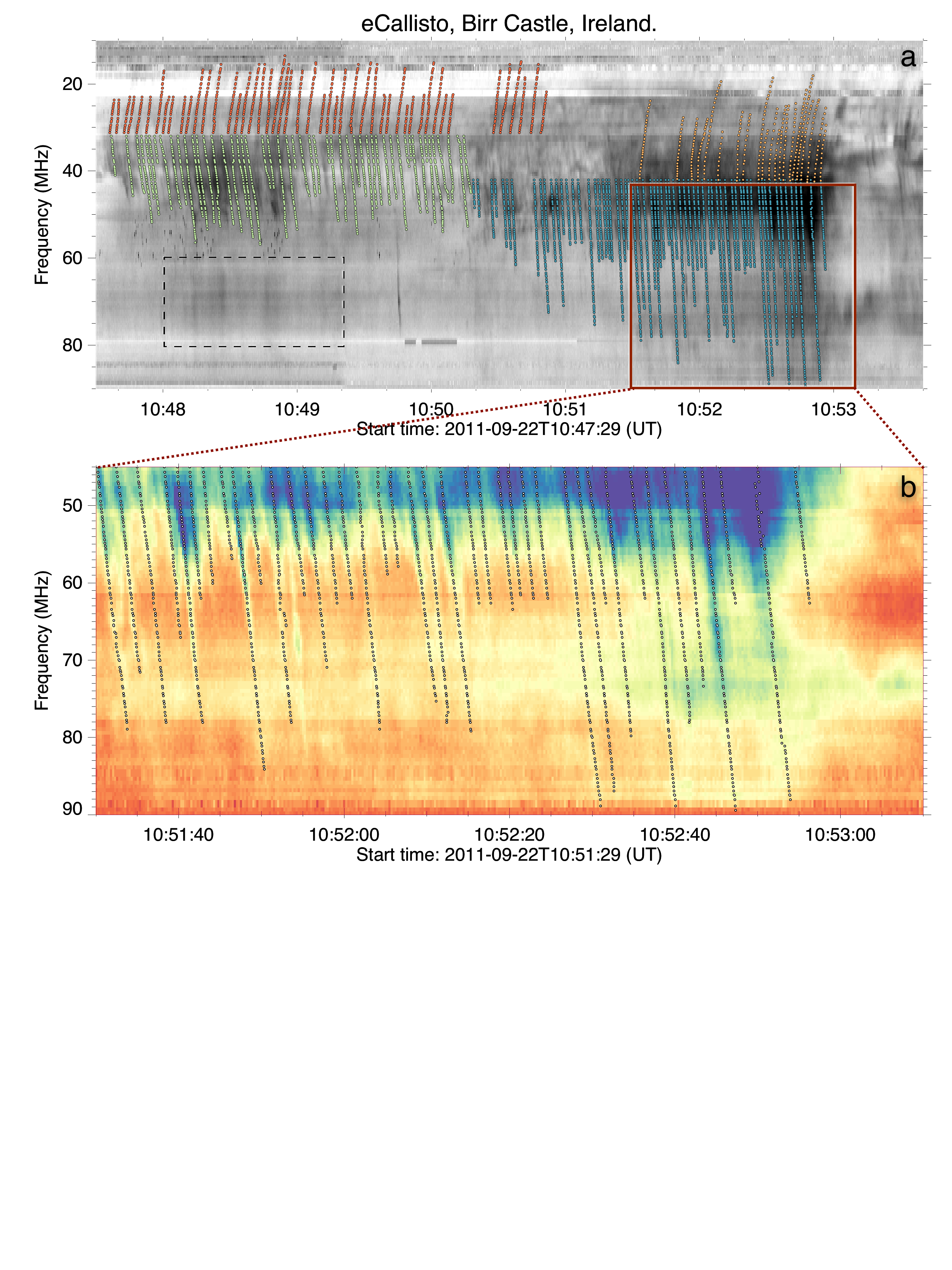}
    \caption[Burst detections.]{(a) Indication of each burst detected in the algorithm using the Hough transform. The forward drift bursts are highlighted in orange-red and reverse drift are in green-blue. In general, the forward drift were more difficult to detect due to their weaker signal to noise and because they drift into an area of high RFI  $<$20\,MHz. This is the reason why there are more reverse drift detections than forward drift. (b) Zoom of the second set of reverse drifting bursts with detections overlaid as circles.}
    \label{fig:detections}
    \end{center}
 \end{figure}

For the herringbone observations, we firstly pass our dynamic spectrum through {\color{black} the \emph{IDL gradient.pro} function to produce an image (dynamic spectrum) of intensity gradient}. This is to produce an edge detection of the radio bursts, delineating the burst peaks more sharply. Next, we choose a section of the dynamic spectrum containing only the reverse drift bursts, such as the dashed box in Figure~\ref{fig:goes_hb}b, and pass it through the Hough transform. Since all bursts in this section have a particular orientation, we know generally which area of Hough space from which to produce a back-projection. This allows us to re-obtain the original image but only containing these bursts specifically, see Figure~\ref{fig:goes_hb}c. Having been passed through the Hough transform and back-projected, the bursts are now sharp, unbroken lines which are much more clearly defined and subject to less RFI or background noise. These sharply defined peak positions allow us to identify the burst peak for each frequency in the dynamic spectrum using a peak finding algorithm. 

{\color{black} We note that the Hough transform fails to demarcate the start and end frequencies of the bursts. In the back-projection, the linear features can continue after the burst has stopped. {\color{black} For example, by comparing Figure~\ref{fig:goes_hb}b and 1c we can see some herringbones drift no higher than 60\,MHz, but in the the back-projection all detections extend as far as 90\,MHz. This is because the transform works by assigning a line to two or more points in the original image (dynamic spectrum), this line is represented mathematically in the transform without bounds. In order for a burst end frequency to be chosen, a user must go through the bursts one by one and select an end frequency i.e. where the burst finally fades into the background emission. This ensures that the correct burst length in frequency is chosen.}

\begin{figure}[t!]
    \begin{center}
    \includegraphics[scale=0.4, trim = 1.7cm 2cm 0cm 1cm]{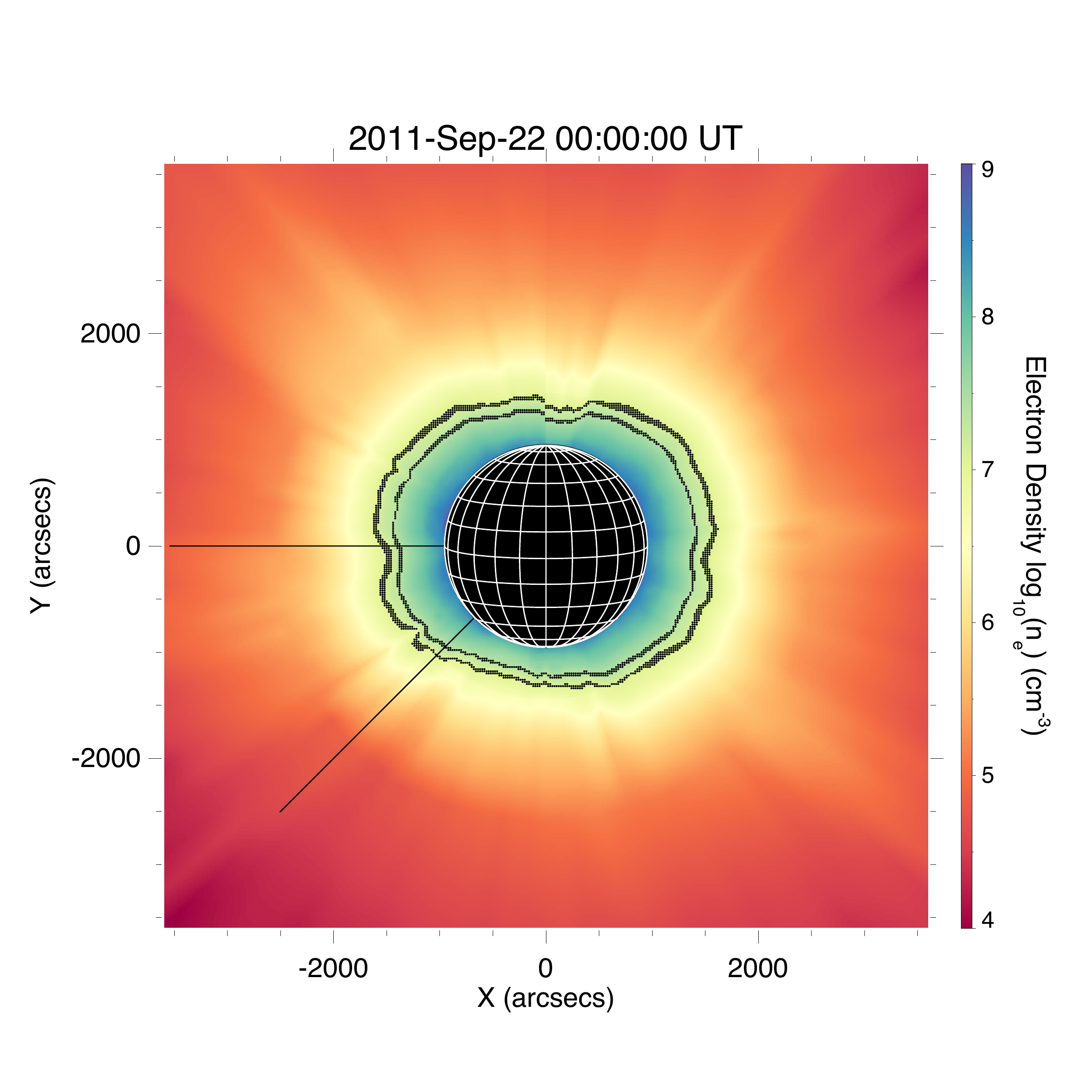}
    \caption[]{{\color{black}Electron density map of the solar corona produced from AIA and LASCO C2 observations \citep{zucca2014}. A density profile of the corona for the data of the event was produced by averaging the density as a function of radius between the lines at position angles of 90$^{\circ}$ and 135$^{\circ}$. This density as a function of radius was used in deriving electron beam kinematics from the herringbone radio burst statistics. The inner and outer black points represent the heights at which 43\,MHz (0.47\,R$_{\odot}$) and 32\,MHz (0.59\,R$_{\odot}$) are expected to occur.}}
    \label{fig:tcd_density_map}
    \end{center}
\end{figure}

The total amount of detections is shown in Figure~\ref{fig:detections}a with forward drift bursts in orange-red and reverse drift in green-blue, with bottom panel showing a zoom for detail. The detection performs well, detecting 188 bursts in total. 
However there are a number of shortcomings. There is not 100\% accuracy, and the transform fails to detect bursts in a particularly noisy background environment, or in regions of low signal to noise. {\color{black} This is particularly notable in a region midway through the activity at 10:50:30\,UT during which there is a shift of the backbone from $\sim$32\,MHz to $\sim$43\,MHz (we discuss the physical interpretation of this in Section 4). During this shift, the activity becomes quite `patchy' and bursts are not so easily detectable, especially the reverse bursts. This is why detections are sporadic around this area.}
Also, the forward drift bursts (those drifting to lower frequencies) are much weaker and generally drift into an environment of high RFI in the AM band below $\sim$20\,MHz, hence detections of forward drift bursts were less successful than reverse drift. In fact, after the backbone shift, the forward drift bursts became very weak, and not easily detectable. These few bursts (yellow detections, Figure~\ref{fig:detections}a) had to be selected by point-and-click. {A test of detection threshold in a quiet area of our dynamic spectrum revealed that detection is successful if the burst has an intensity greater than $\sim$$3\sigma$ above the quiet background, where $\sigma$ is the standard deviation of values in the quiet background. If the burst is surround by noise of comparable intensity (RFI, for example), detection is much more difficult.}

There is a region around 10:48:00--10:49:30 at 60--80\,MHz {\color{black} (dashed box, Figure~\ref{fig:detections}a)} where features were left unselected. They are much fainter and more diffuse than the herringbones, with no discernible drift. We consider these to be possibly part of a continuum emission, so they are excluded here.}

Overall, the algorithm is successful in identifying individual herringbone structure for the majority of time and frequency ranges. It is essentially a semi-automated routine, whereby the transform detects all possible bursts, with the user providing some input to evaluate the validity of the detection. 
 
  \begin{figure}[t!]
    \begin{center}
    \includegraphics[scale=0.35, trim =0cm 0.5cm 5cm 0cm]{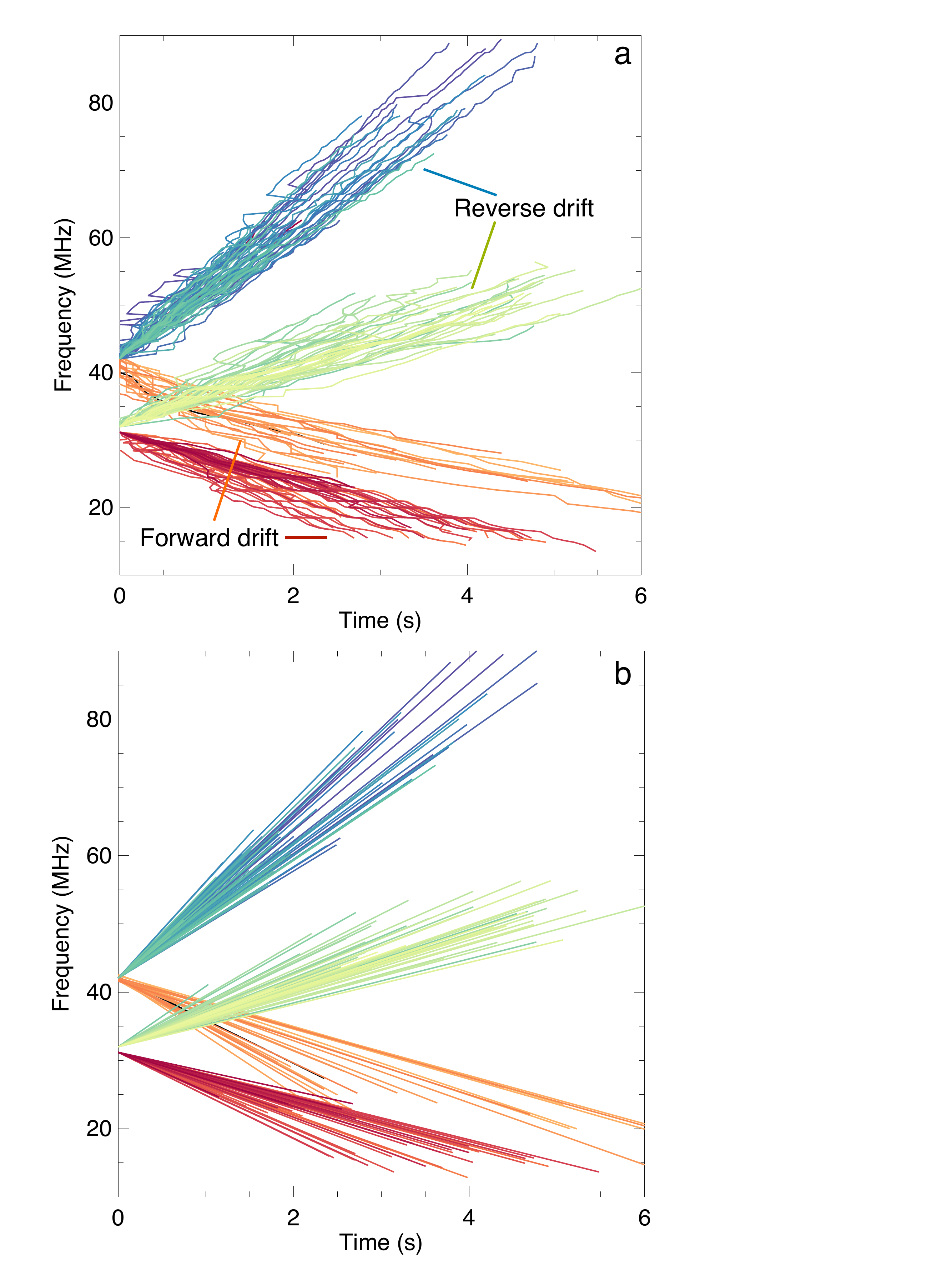}
    \caption[Herringbone frequency time data and fits.]{(a) Frequency vs time data points for all detected bursts for both reverse (blue-green) and forward drifting bursts (orange-red). One line represents one burst, with 188 in total. (b) Linear fits to each burst detection. The green-red lines originate from $\sim$32\,MHz and are detections of the first set of bursts between 10:47:30--10:50:30\,UT. The blue-orange lines are the second set of bursts originating from $\sim$43\,MHz between 10:50:20--10:53:00\,UT. There are not equal numbers of positive and reverse drift bursts, indicating there is not a one-to-one correspondence of particle beams travelling toward and away from the sun, i.e. beams are bi-directional but not always in reverse and forward drift partners.}
    \label{fig:ft_data_fits}
    \end{center}
\end{figure}


\section{Results}

In the following we refer to the herringbones as observed in dynamic spectra as bursts, while the population of fast electrons causing the radio burst are called a `beam' i.e. one beam of electrons causes one herringbone burst.

The detection of 188 bursts allows for the statistical analysis of some useful properties. The total number of reverse and forward drift bursts are 118 and 70, respectively. It is interesting to note that there is not a reverse drift for every forward drift burst and vice versa. This shows that the beams are bi-directional in space, but do not always emerge from the shock as sunward and anti-sunward partners (as noted in \citet{cairns1987}) -- we discuss this further in Section 5. 

{\color{black}In the following we derive kinematical results of the particle beams causing the herringbones. To do this we make the assumption that we observe fundamental plasma emission. It is reasonable to assume that we observe the fundamental component of the burst, given that fundamental flux exceeds harmonic flux in 80\% of herringbone observations \citep{cairns1987}. However, as we discuss below the general results for beam kinematics do not change if we assume harmonic emission.}

\subsection{Burst altitude and backbone shift}

It is apparent from Figure~\ref{fig:detections}a that mid-way through the radio activity at $\sim$10:50:30\,UT, there is a shift of the backbone to higher frequency, from $\sim$32\,MHz to $\sim$43\,MHz. This could indicate the acceleration region shifted in altitude midway through the radio activity. To calculate this altitude shift we firstly assume fundamental plasma emission ($f_{plasma}$) and convert to electron number density ($n_e$) via $f_{plasma}\sim8980\sqrt{n_{e}}$. The height at which this density occurs was found using a TCD density model \citep{zucca2014}, produced from density measurements of the corona using the Atmospheric Imaging Assembly \citep[AIA;][]{lemen2012} and Large Angle Spectrometric Coronagraph \citep[LASCO;][]{bru95} C2 observations on the date of the event, see {\color{black} Figure~\ref{fig:tcd_density_map}. An average of the density profile was taken between position angles of $90^{\circ}$ and $135^{\circ}$}, i.e. where this event took place off the limb (see ~\citet{carley2013}). This density profile (produced from observations) prevents the needs for a `guess' model, and allows us to be more confident that the density model used in the analysis is a correct description of the corona for this event. We find that the acceleration region is initially at constant altitude of 0.59\,$R_{\odot}$ then undergoes a significant shift to a lower altitude of 0.47\,$R_{\odot}$. 

Although a shift in frequency is generally taken as an altitude shift, we cannot rule out that a shift in frequency may also be from an accelerator at a constant altitude encountering a change in background density profile. For example, if the accelerator propagated at constant altitude, but changed in position angle it may encounter larger densities, resulting in a shift to higher emission frequencies for the herringbones. Figure~\ref{fig:tcd_density_map} shows that different density profiles exist within the bounds of where the event took place. We stress that the shift in frequency is due to an
accelerator moving in an inhomogenous corona, either shifting in height or position angle (or both). Either way, it is clear that the accelerator changes environment midway through the burst. We show later in Section 4.3 that this results in a change in burst intensity profile and speed.

{It is also worth noting the density calculated from the TCD model is produced from a line-of-sight average. This inherently ignores the spatial variability in the corona, for example at the interface between open and closed field. The actual densities may be higher or lower than the average we have calculated. 
However, \citet{zucca2014} showed that for this event the expected heights we calculate from the density map are a close match to the heights of radio sources observed by NRH, confirming that our density values are reliable.}

%
%
\subsection{Burst drift rates and particle beam speeds}

\begin{figure}[t!]
    \begin{center}
    \includegraphics[scale=0.38, trim=0cm 1cm 1cm 0cm]{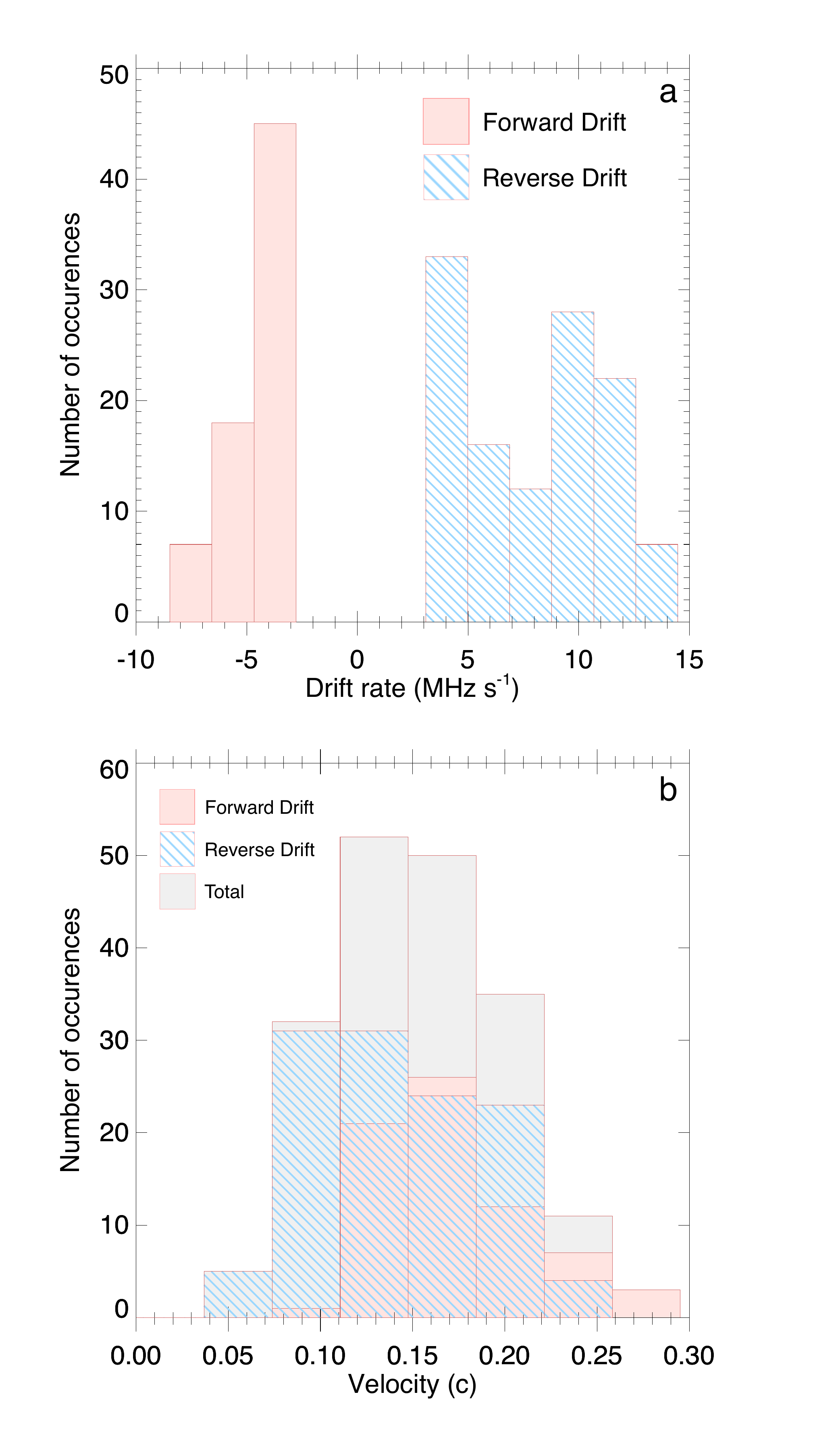}
    \caption[]{(a) Histogram of drift rates of reverse drift (blue stripes) and forward drift (red) bursts. 
    The reverse drift bursts have a bimodal distribution. Those with a generally higher drift rate (above 10\,MHz\,s$^{-1}$) are the reverse drifting bursts that start at 43\,MHz. Taking the absolute value of these drifts and using a TCD density model we produce a histogram of velocities for the electron beams causing the herringbones (b), with an mean of 0.16\,c. We note that although the drift histogram is bimodal, the velocity histogram is not. This is due to electron beams of a single velocity distribution propagating in different heights of the atmosphere. A different height means a different density gradient and hence a different drift rate.}
    \label{fig:drift_vel_hist}
    \end{center}
\end{figure}

Figure~\ref{fig:ft_data_fits} shows the frequency time traces of all detected bursts (a), along with linear fits to the data (b). These linear fits were used to obtain the drift rates of all herringbones in the radio burst and produce a histogram, as shown in Figure~\ref{fig:drift_vel_hist}(a). The bin size is calculated from the smallest shift detectable in the dynamic spectrum, taken to be two pixels in frequency (0.9\,MHz) and two pixels in time (0.5 seconds) giving a minimum detectable drift of 1.8\,MHz\,s$^{-1}$. The histogram of burst drift rates were also converted to beam velocities using the TCD density model Figure~\ref{fig:drift_vel_hist}(b). The bin size for velocity is the minimum detectable drift rate converted to velocity using our density model (0.04\,c).

\begin{figure}[t!]
    \begin{center}
    \includegraphics[scale=0.27, trim= 1cm 1cm 0cm 0cm]{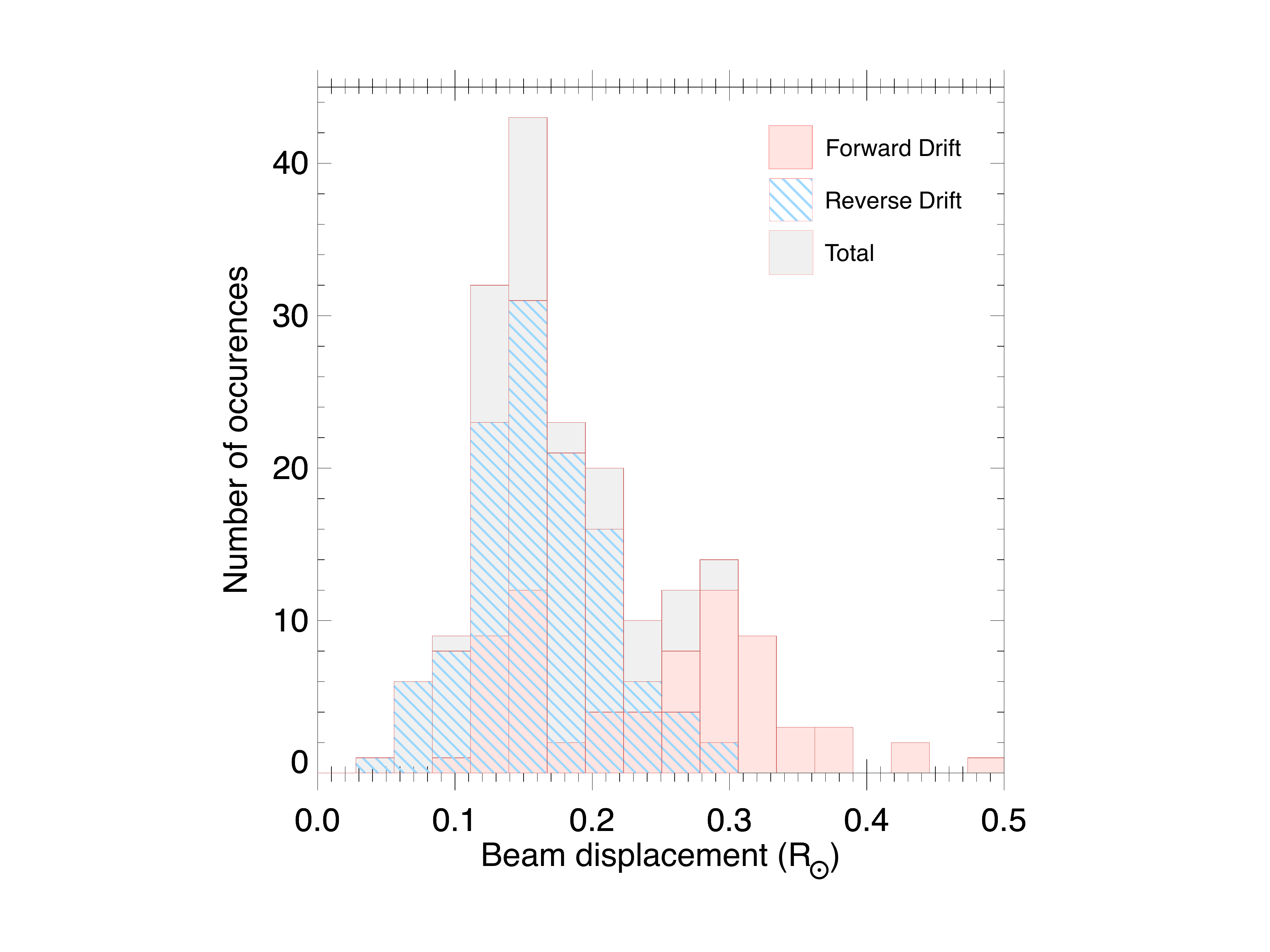}
    \caption[]{Histogram of beam displacements, produced by converting the frequency span of individual bursts in the dynamic spectra to distances using a TCD density model. An interesting result here is that the forward drift bursts (those travelling away from the sun) tend to travel larger distances than the reverse drift (those travelling toward the sun). It is clear that the positive drift bursts solely make up the high-end tail of this histogram. This may be an indication that electron beams travelling toward the surface stop travelling (or stop emitting) sooner than those travelling away from the sun. The bin size is calculated from minimum detectable shift over two pixels (0.9\,MHz) in the dynamic spectrum, giving a 0.03\,$R_{\odot}$ bin size.}
    \label{fig:displ_hist}
    \end{center}
\end{figure}

The drift rate histogram has some notable features. Forward (red) and reverse (blue stripes) drifts have similar absolute magnitudes, however the reverse drift bursts appear to be bimodal. The high drift population (greater than $\sim$10\,MHz\,s$^{-1}$) belong to the reverse drift bursts which start at 43\,MHz (between 10:50:20--10:53:00\,UT). This is not necessarily because the beams causing these bursts have a higher velocity, but because they are accelerated lower in the atmosphere,  and therefore encounter a steeper gradient in density, giving them a steeper drift in frequency space. This can be confirmed by analysing the velocity histogram in Figure~\ref{fig:drift_vel_hist}(b). {\color{black} Forward and reverse beams belong to similar and overlapping velocity distributions, with their total histogram} peaked at $\sim$0.16\,c (maximum and minimum speeds of 0.36 and 0.06\,c, respectively, similar to those reported for herringbones {\color{black} at $\sim$10--120\,MHz} by \citet{cairns1987}), with the beams producing forward and reverse drift bursts showing only a slight difference in speed distribution. This is a good indication that we have detected drift rates accurately, i.e. independent drift rates and varying start frequencies result in similar velocities for all electron beams. This is evidence that both forward and reverse beams throughout the entire radio burst belong to the same accelerator even though there is an altitude shift halfway through the activity. {\color{black}The effect of assuming harmonic emission does not change this result, it simply shifts the velocity histogram to higher values by $\sim$$0.05$\,c. 

\begin{center}
\begin{table*} 
\centering   
  \caption{{Compilation of radio burst properties and derived particle beam kinematics using a TCD density model. F is forward drift and away from the sun, while R is reverse drift and toward the sun.}}

    \begin{tabular}{lcccccc}
        \hline
        	     & \# 		& Drift (MHz\,s$^{-1}$) 		& Time (s) 		& Vel. (c) 		& Displ. (Mm) 		&  		Stop altitude  (R$_{\odot}$) \\
		     &  		   &  min \hspace{0.3mm} mean \hspace{0.3mm} max 		&  mean  \hspace{0.3mm}  mode 		& min \hspace{0.3mm} mean \hspace{0.3mm} max 		&  min \hspace{0.3mm} mean \hspace{0.3mm} max  		&  min \hspace{0.3mm} mean \hspace{0.3mm} max  \\
		 \hline				
		        R & 118 	&   3.0  \hspace{0.3mm} 8.1   \hspace{0.3mm} 14 	&  2.8 \hspace{0.3mm} 2.1  	&  0.06  \hspace{0.3mm} 0.14 \hspace{0.3mm}  0.25 		&  36 \hspace{0.3mm} 112 \hspace{0.3mm} 197 		&  0.2 \hspace{0.3mm} 0.34  \hspace{0.3mm}0.45 \\
		F &  70 		&   -8.5  \hspace{0.3mm} -4.6 \hspace{0.3mm} -2.8 		&  3.3 \hspace{0.3mm} 2.6  		&  0.11  \hspace{0.3mm} 0.18  \hspace{0.3mm} 0.27 		&  \hspace{0.4mm}  74 \hspace{0.3mm} 170 \hspace{0.3mm} 344  & - \\
	    \hline
	    \end{tabular}
        \label{table1}
   
\end{table*}
\end{center}

\subsection{Beam displacements}

Figure~\ref{fig:displ_hist} shows a histogram of electron beam displacements {\color{black}(distance that the electron beam travels)} in space. The displacements were calculated by converting start and end frequency to start and end height using our density model. The bin size is calculated from the minimum detectable frequency shift in the dynamics spectrum (2\,pixels = 0.9\,MHz) converted to a displacement from the model (0.03\,$R_{\odot}$). The overall distribution appears bimodal, with 
the upward electron beams making up the high-end tail of the distribution (greater than 0.3\,$R_{\odot}$). {\color{black}} 

This is an interesting result, showing that the beams travelling toward the solar surface are either stopped sooner or stop emitting sooner. Sunward beams travel in the range of 36--196\,Mm with an average of 112\,Mm, while those travelling away from the sun have a larger displacement in the range of 73--344\,Mm with an average of 170\,Mm. {\color{black}In fact, the upward-beam displacements may be longer than this; the stopping frequency of forward drifting bursts is more difficult to identify, as the bursts generally drift into a region of high RFI and below $\sim$20\,MHz toward the ionospheric cutoff frequency.}  {\color{black} This is the reason for the bi-modality of the forward drift bursts. Those forward drift bursts that belong to the small displacements in the distribution ($\sim$0.15\,$R_{\odot}$ in the red histogram) primarily belong to the first set of forward drifters in Figure~\ref{fig:detections}a. It can be seen here that some of these detections (red circles) are cut short because they drift into a region of RFI, just below 30\,MHz. These bursts likely drift to lower frequencies and hence should have larger displacements.} The reason for upward beams travelling further (or emitting for longer) than sunward beams is discussed in the Section 5.

Since the velocity of both reverse and forward drift particles are the same, the lifetime of the forward particles should be slightly longer, given that their displacements are longer. We confirm this by looking at the mean and mode lifetime of the bursts in Table 1, which shows that the forward particles generally have a longer lifetime by $\sim$0.5\,second, i.e. travelling for 0.5 extra second at speeds up to $\sim$0.27\,c gives an extra distance of $\sim$40\,Mm, which is similar to the extra displacement range we find for the forward bursts.

{\color{black}Again, the effect of assuming harmonic emission does not change the result. Both forward and reverse drift beams are shifted to smaller displacements by $\sim$$15$\,Mm, however the forward drift beams still make up the high-end tail of the displacement histogram and on average travel a further distance.} 
{The radio burst and derived electron beam kinematic properties, as well as the difference between forward and reverse drifters, are listed in Table 1.} 

{Finally, we tested all kinematics results using the popular Saito, Leblanc, Newkirk, Mann, and Baumbach-Allen density models \citep{saito1977, leblanc1998, newkirk1961, mann1999, allen1947}, see Table~\ref{table2}. Although the absolute values of the derived kinematics differ slightly, the general outcome is the same for all models. Table~\ref{table2} also serves to highlight the variable results one might expect when using these density models to derive radio burst kinematics. Here we take the TCD model to be the most reliable, as it is based on density estimates from the day of the event. We reiterate that the TCD density model provides a reliable description of density values in the corona, given the close agreement of NRH observed source heights and the TCD model heights \citep{zucca2014}.}}

\begin{center}
\begin{table} [b!]
\centering   
  \caption{{Comparison of beam kinematics for different density models. All quoted values are mean values. The beam displacements are shown for forward (F) and reverse (R) drifters separately.
  			As expected, the values differ among the different models. However, the general results are unaffected by a change of density model.  \newline
			{\tiny ${\dagger}$ Only valid if the shift in frequency of the backbone is interpreted as an actual source altitude shift.} \newline
			{\tiny ${*}$ Baumbach-Allen.}}}

    \begin{tabular}{lccc}
        \hline
        	       {\tiny Model} &    {\tiny Altitude Shift}$~{\dagger}$  {\tiny (R$_{\odot}$)} &   {\tiny Beam Displ. (Mm) }	  & {\tiny Beam Speed (c)} \\
	      					           	&      {\tiny start $\rightarrow$ end    }                      &  F \hspace{8mm}R                   &  F \hspace{8mm}R   \\
	  \hline   
			TCD  &  0.59 $\rightarrow$ 0.47	  &  170	 \hspace{5mm} 112 &  0.18  \hspace{5mm} 0.14  \\
			Saito 	&    0.53 $\rightarrow$ 0.39	  &  218 \hspace{5mm} 132	 &  0.22  \hspace{5mm} 0.17  \\
			Leblanc &  0.38 $\rightarrow$ 0.25	  &  202 \hspace{5mm} 123	 &  0.2  \hspace{5mm} 0.16  \\
			Newirk  &     0.74 $\rightarrow$	0.57  &  286 \hspace{5mm} 141 &  0.28  \hspace{5mm} 0.18  \\
			Mann 	&    0.35 $\rightarrow$ 0.25  & \hspace{-2mm} 165 \hspace{6mm} 90 &  0.26  \hspace{5mm} 0.17	 \\
			B-A$^{*}$	 &   0.56 $\rightarrow$ 0.41	  &  261 \hspace{5mm} 128	 &  0.16  \hspace{5mm} 0.12  \\
	  \hline
	    \end{tabular}
        \label{table2}
   
\end{table}
\end{center}

\subsection{Burst intensity correlations}

\begin{figure}[t!]
    \begin{center}
    \includegraphics[scale=0.4, trim=0cm 1cm 4cm 1cm]{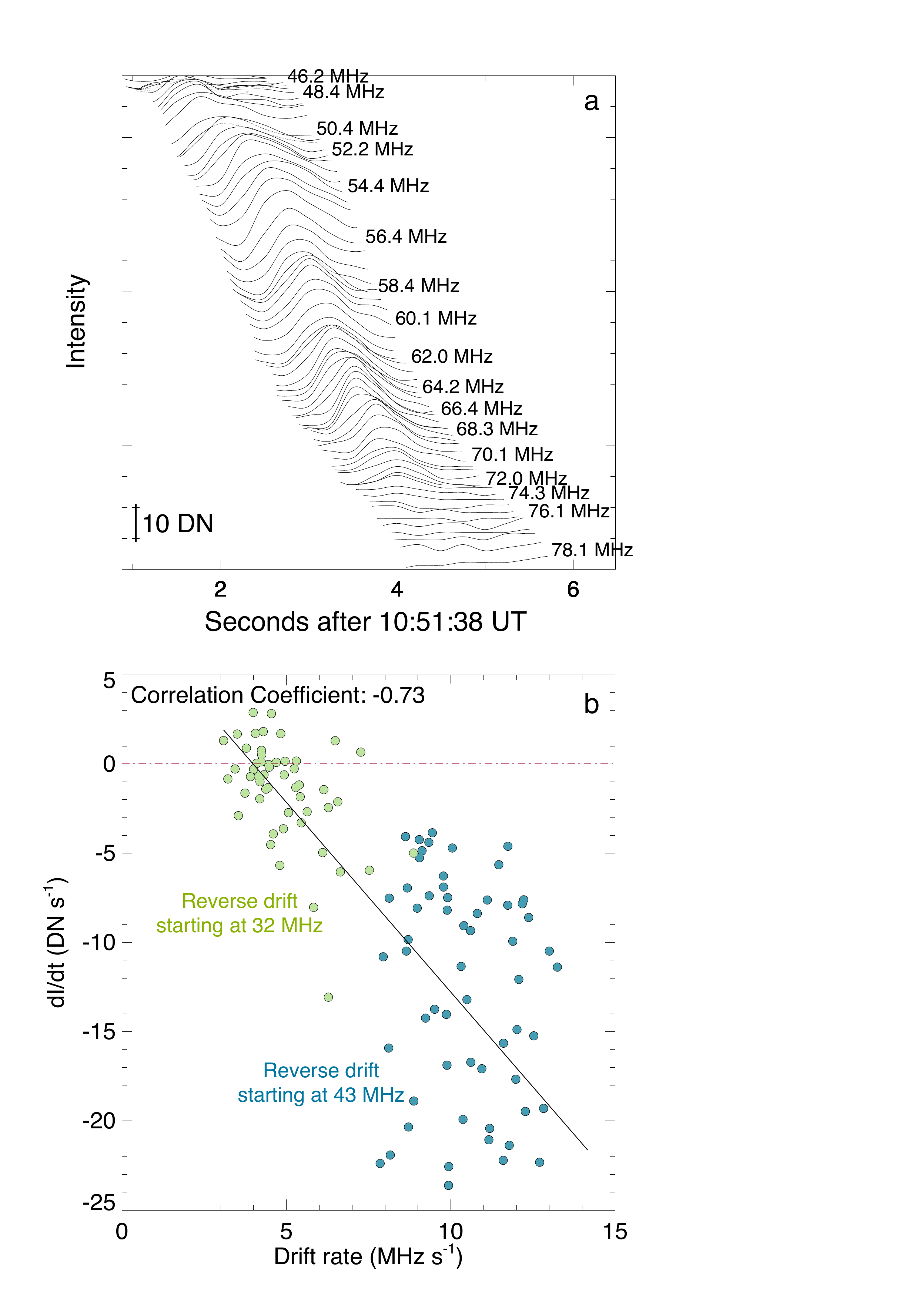}
    \caption[]{(a) Example of a single herringbone burst intensity at each frequency. The burst is most intense at its start and decreases in intensity as it drifts to higher frequencies, i.e. decreases in intensity over its lifetime. (b) Scatter plot of burst drift rate vs the rate of decrease in intensity along the burst for the reverse drift bursts only. The receiving system is uncalibrated, so the intensity are simply quoted in Data Numbers or (DNs). The green points are the reverse drift bursts starting at 32\,MHz ($R_{32}$), while the blue points are reverse drift bursts starting at 43\,MHz ($R_{43}$). We note that the $R_{32}$ have a lower drift and rate of change of intensity than the $R_{43}$ bursts. Overall, a clear negative correlation is shown here; this occurs because the majority of the bursts decrease in intensity over their lifetime, i.e. the slope of the black line is $(dI/dt) = m (df/dt) \Rightarrow m= dI/dt \times dt/df=dI/df$. This is the opposite of type III bursts, which generally show an increase in intensity over their lifetime.}
    \label{fig:scatter_drifts}
    \end{center}
\end{figure}

In the following we analyse the correlations amongst burst intensity and other parameters, as well as the bursts rate of change of intensity over the burst lifetime. Figure~\ref{fig:scatter_drifts}(a) gives a single example of how a burst maximum intensity changes over its lifetime, or equivalently, changes with respect to frequency. In Figure~\ref{fig:scatter_drifts}(b), we plot burst drift rate vs the rate of change of intensity over the burst lifetime. This plot contains only the reverse drift bursts. This was done because the reverse drift bursts have a much better signal to noise, are subject to much less RFI and their intensity can generally be sampled for most of their duration -- this is not the case for the forward drift bursts. 

The system at RSTO is uncalibrated, so intensity is quoted in data numbers (DN). We can only note that the receiver response is logarithmic in intensity. {\color{black} The data have also been background subtracted using the \emph{SolarSoft constbacksub.pro} routine. This finds a relatively quiet spectra of the spectrogram (containing no bursts) and subtracts the quiet spectra from each time step in the spectrogram. Since the RSTO Callisto system has a logarithmic response to any signal \citep{benz2005}, a subtraction of a background in this manner is essentially a division of the spectrogram by the background. This flattens any variation in intensity that the system may have across the bandwidth, so changes in intensity due to instrumental response may be ruled out. Furthermore, lack of knowledge of the absolute flux values should not affect our general results. }

Figure~\ref{fig:scatter_drifts}(b) shows an overall negative trend with a correlation coefficient of -0.73. The negative trend reflects the fact that the bursts decrease in intensity over time, with the slope of the fit to the data being equal to $dI/df$ i.e, the rate of change of intensity over the bursts' duration in frequency. {\color{black} Decreasing intensity as the burst drifts in the dynamic spectrum} is the opposite to what is found in type III bursts \citep{cairns1987, Saint-Hilaire2013}, meaning their may be a fundamental difference to the propagation characteristics of type IIIs and herringbones, or the environment in which the electron beams propagate may be different. 

Figure~\ref{fig:scatter_drifts} also shows the separation of the distribution into two populations, those belonging to bursts that start at 32\,MHz (green circles) and those that start at 43\,MHz (blue circles). The green population has both smaller drift and smaller change in intensity, with some having an \emph{increase} in intensity over time. Four of the bursts from the green population show a slow rise and fall of intensity with time.  

The different morphological characteristics of the 32\,MHz bursts may be due to the altitude and/or coronal environment shift of the accelerator as described above. Once the acceleration region shifts to a new environment, the radio bursts increase both in drift rate and the rate of change of intensity (as seen from the change from green to blue points in Figure~\ref{fig:scatter_drifts}). This suggests that the appearance of herringbones may depend on the kind of coronal environment in which the electron beams propagate.

Finally, we investigate any correlation between intensity and beam displacement and velocity for the reverse drift bursts, shown in Figure~\ref{fig:displ_maxi}(a). The top figure shows beam displacement in space vs the maximum intensity of the bursts above the local background. The overall correlation is positive but weak, with a correlation coefficient of 0.48. However, the two populations of reverse drift bursts again show a difference. Those with start frequency of 43\,MHz (blue) are generally more intense and travel further than those starting at 32\,MHz (green). The separate populations by themselves exhibit positive correlations (0.6 and 0.55), showing the more intense the radio burst, the further the beam travels in space (toward the solar surface). We also test for any correlation between burst intensity and beam velocity shown in Figure~\ref{fig:displ_maxi}(b); we find {\color{black} almost no correlation} here with correlation coefficient of 0.31. However we do note that the two populations of reverse drift bursts are separated, with the faster bursts generally being more intense on average. {\color{black}Again, the assumption of harmonic emission does not affect the above result.}


\section{Discussion}

In this section we discuss the results in terms of the coronal environment of the herringbones and with regard to the standard theories of radio burst intensity, i.e. wave mode conversion from Langmuir waves to ion acoustic and electromagnetic waves {\color{black} \citep{robinson1992, robinson1994, li2012, ratcliffe2014}}.

\subsection{Burst starting frequencies}

An interesting feature of these herringbones is that both forward and reverse drift bursts originate at the same frequency (Figure~\ref{fig:detections}(a)). {\color{black} This may give clues as to the shock region in which the beam propagates and/or the beam formation and emission mechanisms at play.}

{\color{black} Firstly, the reverse and forward drifters having the same start frequency} indicates that electron beams travelling toward and away from the sun begin propagating in a similar density environment. This may be evidence that the forward and reverse drift bursts are not from beams ahead and behind the shock, respectively. If this were so, the reverse drift bursts would start at a separate and higher frequency because of reverse electron beams encountering a sudden downstream density jump across the shock. Instead we propose that forward and reverse drift bursts are both from electron beams accelerated upstream of the shock. For example, this would occur if a wavy shock propagated laterally in the corona and constantly accelerated electron beams onto field lines in its upstream region, such as that illustrated in Figure 5 of \citet{zlobec1993} or in the illustration of \citet{cliver2013}.

{\color{black} Secondly, we may interpret the similar start frequencies by comparing herringbones to type III observation and theory}. The starting frequency of type III bursts depends upon a number of initial electron beam characteristics, namely the injection height, size, time and energy distribution of the beam \citep{kane1982, reid2011, reid2014, reid2014a}.  The beam has to travel a certain distance before velocity dispersion will create an unstable electron distribution, assuming the injection profile is a power-law.  We do not observe any separation in the starting frequencies between the forward and reverse herringbone bursts.  A lack of separation can be explained with very fast ($<0.1$\, s) injection times in an accelerator that is small along the dimension of electron propagation.  This scenario could be feasible for electron acceleration originating in a small diffusive region in a shock.  Another explanation is that the injected electron distribution is initially unstable to Langmuir waves. Essentially the accelerator creates a pre-formed beam at some velocity $v>>v_{T_e}$ (where $v_{T_e}$ is the thermal electron temperature), leading to bi-directional bursts with no frequency gap. Such characteristics may be an indicator of the acceleration process at play in the shock.

{\color{black} We stress that it may be upstream beam propagation \emph{and/or} immediate beam formation that may lead to the same start frequency of reverse and forward drifters.}

\begin{figure}[t!]
    \begin{center}
    \includegraphics[scale=0.45]{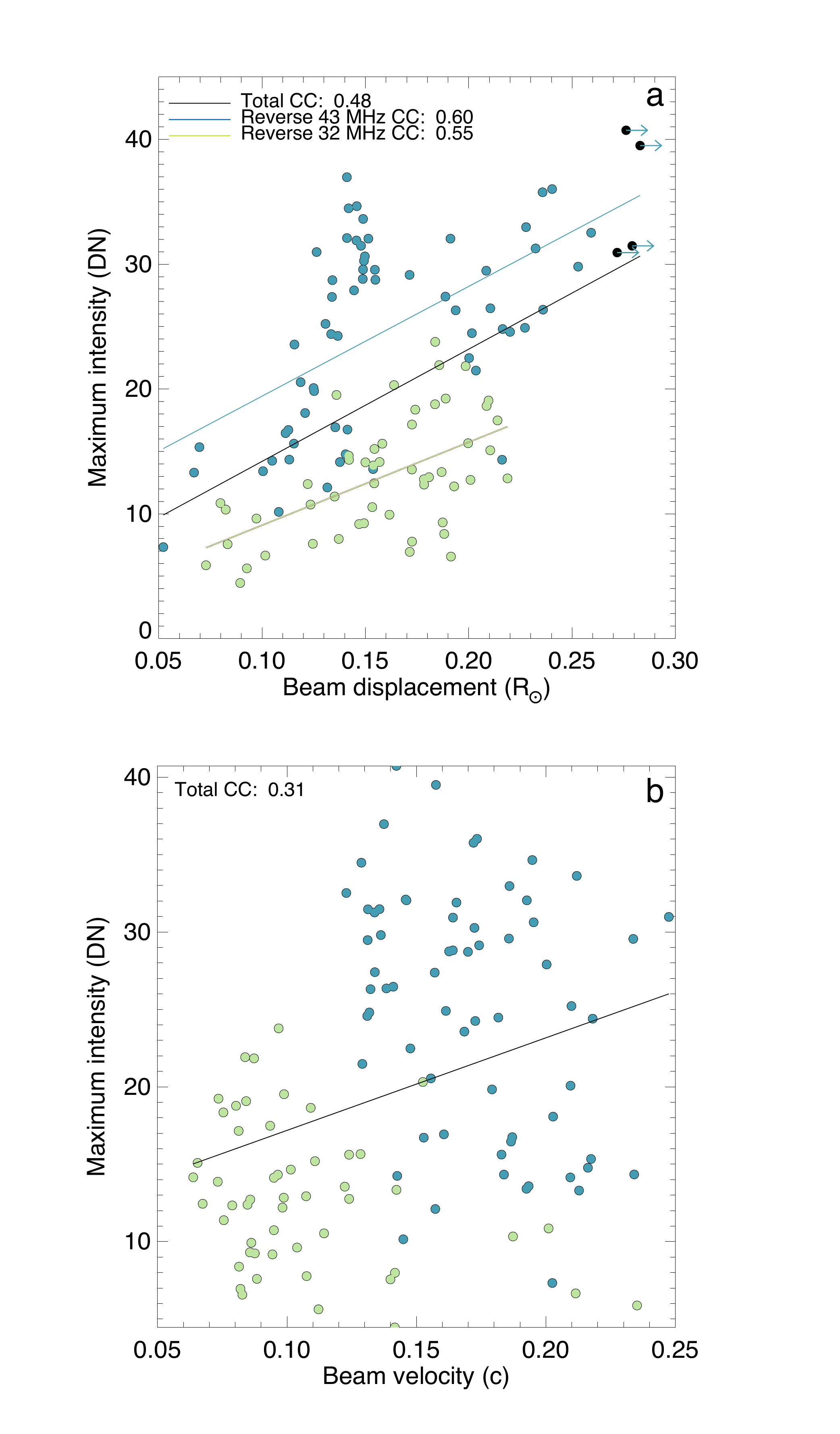}
    \caption[]{(a) Scatter plot of beam displacement (using the TCD density model) against burst maximum intensity above the background, for reverse drift burst only. All bursts together show a weak positive correlation coefficient of 0.48, meaning the more intense the bursts the further they may travel in space. There is a clear separation in intensity between the two types of reverse drift bursts. By themselves, the reverse drift bursts starting at 43\,MHz have a correlation coefficient (CC) of 0.60, while those starting at 32\,MHz have CC of 0.55. The black circles with arrows indicate that these bursts may be slightly longer in space. Their corresponding bursts drifted outside the dynamic spectra observing range. (b) Maximum intensity vs beam velocity, there is almost no correlation here (CC$\,=0.31$), with the only dependency being with the two different sets of bursts, i.e. one set (green points -- start frequency of 32\,MHz) are less intense and slower than the other.}
    \label{fig:displ_maxi}
    \end{center}
\end{figure}

\subsection{Displacements and emission mechanisms}

Although they have similar velocities, the forward and reverse drift bursts show a significant difference in the distances the electron beams travel. The anti-sunward beams travel a distance of 170$_{-97}^{+174}$\,Mm away from the acceleration site (and possibly further, given the herringbones may reach frequencies below our observational limit at $\sim$15\,MHz), while those travelling toward the sun come to a stop sooner, reaching a smaller distance of 112$_{-76}^{+84}$\,Mm. The sunward beams either stop travelling and/or stop emitting possibly due to them encountering a different environment the lower in the corona they reach. The reason for this may be in the emission/absorption mechanisms at play in the corona, which has been investigated for type III bursts.

Observationally, reverse type III bursts are much less frequently observed than standard type III bursts and are observed at much higher frequencies during flares \citep[e.g.][]{melendez1999}.  This is due to a number of reasons: difficulty of the radio waves to escape the low corona due to absorption, an increase in the background electron density, and the presence of a positive background electron density gradient. Furthermore, it is harder to excite Langmuir waves in denser plasma.  The growth rate of Langmuir waves is inversely proportional (amongst other terms) to background density or $\gamma_L \propto n_e^{-0.5}$.  
In a higher density environment, Langmuir waves are refracted out of resonance with the electron beam to higher phase velocities (lower values in wave number space or \emph{k}-space) where a smaller flux of electrons is present \citep[see e.g.][]{kontar2001, reid2013, krafft2013}.  This results in fewer Langmuir waves to convert to radio waves and may explain the reason why the reverse drifters stop emitting sooner. {As an example of this, bi-directional type III bursts at high frequencies (in the metric/decimetric range) have been simulated by \citet{li2008, li2011}. They highlight the enhanced difficulty that an electron beam has in producing radio emission with a positive drift rate.}

{A second factor at play here could be the collisional losses experienced by the electron beams. To test this assertion, we were able to estimate the column densities encountered by the electron beams from $\int n(r)_{tcd}dr$, where $dr$ is integrated along the path length of the beam and $n(r)_{tcd}$ is from our density model. We find that the anti-sunward beams encounter a column density of $1.4\times10^{17}$\,cm$^{-2}$, while the sunward beams encounter $1.1\times10^{18}$\,cm$^{-2}$. The lower column densities for the outward propagating beams may not be sufficient to significantly damp the beams, especially for the beam speeds we observe in Figure 5. However, the average column densities for the downward propagating beams are higher ($1.1\times10^{18}$\,cm$^{-2}$), enough to collisionally stop electrons at 3 keV (0.1c) or less \citep{brown1973, emslie1984}. Given the sunward beam speeds of 0.14c, it is possible that Coulomb collisional energy losses play a role in stopping these beams. However, it is not clear whether collisions are the dominant energy loss mechanism that causes radio emission to cease.  We observe many drift rates with higher velocities in Figure 5 and we do not observe a correlation between inferred drift rate velocities and inferred beam displacement. We simply highlight collisional losses due to higher column densities as a possible factor in the shorter displacements of the sunward beams.}

\subsection{Intensity, velocity, and number density}

We also investigated the dependency of burst intensity on other characteristics. A number of theoretical studies have been performed on radio burst emissivity {\color{black}\citep{melrose1986, robinson1992, robinson1993, robinson1994, li2012, ratcliffe2014}}. The general process of plasma emission involves the formation of a particle beam through some acceleration process; the plasma becomes unstable due to the presence of this beam, resulting in the growth of Langmuir waves; these Langmuir waves 
coalesce with ion acoustic waves or secondary Langmuir waves to produce electromagnetic emission at the local plasma frequency and its harmonic. To compute the expected emissivity of the resulting electromagnetic (radio) waves, stochastic growth theory has been employed in a number of cases to successfully explain the observed intensities of radio bursts \citep{robinson1993a, knock2003, schmidt2012}. The theory implies that burst emissivity should depend on the beam velocity and number density, given by
\begin{equation}
\centering
    j_M(r) \approx \frac{\Phi_M}{\Delta\Omega_M}\frac{n_b m_e v_b^3}{3l(r)}\frac{\Delta v_b}{v_b} ~;
    \label{eqn:plasma_emiss}
\end{equation}
here the $M$ stands for either fundamental $F$ or harmonic $H$ emission, $\Delta\Omega_M$ is the solid angle over which the  emission is spread, $n_b$ is the electron beam number density, $v_b$ is the beam speed, $l(r)$ is the distance from emission point to observer, $\Delta v_b$ is the width of the beam in velocity space, and $\Phi_{F, M}$ are wave conversion efficiency factors which depend on $v_b$.

However, we find only a weak correlation between burst intensity and velocity (Figure~\ref{fig:displ_maxi}b). The weak correlation could be related to variations in the beam density; a parameter that we are unable to infer.  If the accelerator energises electron beams with a weak correlation between beam density and beam velocity then we would not expect a tight correlation between burst intensity and beam velocity.  Moreover, the presence of density fluctuations, known for modulating the type III burst intensity \citep[e.g.][]{muschietti1985, reid2010, ziebell2011, li2012} can increase the scatter between beam velocity and maximum intensity. 

The acceleration region produces electron beams with more intensity when the starting frequency shifts to 43 MHz.  Average values for the maximum intensity, beam displacement and beam velocity all increase (Figure 8) for the reverse drift herringbone bursts.  Figure 8 suggests that the faster beams that produce higher intensity bursts propagate further toward the solar surface to reach higher frequency plasma.  This is similar to the results of \citet{cairns1987}, in which they showed that bursts with greater velocity reached further distances from the acceleration site.  Interplanetary type IIIs display this property with \citet{leblanc1995, leblanc1996} finding lower stopping frequencies for type III bursts with higher radio flux. Recent work by \citet{reid2015} found that beam density and injected spectral index both affect the stopping frequency of type III bursts.  Overall, Figure~\ref{fig:displ_maxi} would suggest that beams that are greater in number density and/or velocity travel a larger distance in the corona.

\subsection{Accelerator environment change}

Midway through the radio activity, the origin of acceleration either drops significantly in altitude from 0.59\,R$_{\odot}$ to 0.47\,R$_{\odot}$ or experiences an environment change. This is detectable through the increase in start frequency from 32\,MHz to 43\,MHz of both forward and reverse drift bursts. The shift occurs at around 10:50:30\,UT and a change in the characteristics in herringbones can be seen. Firstly, the reverse drift bursts are initially morphologically different in the dynamic spectrum, with some bursts either showing an increase in intensity over time and others showing an increase then decrease. After the shift, the herringbones acquire a faster drift rate and larger (more negative) rate of change of intensity. 
Figure~\ref{fig:scatter_drifts} shows this clearly, and confirms that there is a morphological difference in the herringbones before and after the shift in altitude. This implies that the presence of herringbones in a radio shock signature could have a dependency on the environment in which the shock propagates. This is in addition to the theories of a wavy shock front producing herringbones \citep{zlobec1993, vandas2011}, i.e. herringbone formation may depend as much on coronal environment as it does on shock characteristics, which has not been shown in the past.

\subsection{Alternative acceleration mechanisms?}

Finally, herringbones have traditionally been interpreted as being formed by a shock and there is much evidence to suggest a shock origin \citep{cairns1987, cane1989}. Various models have been produced to explain the existence of herringbones, such as a wavy shock front on a CME flank \citep{zlobec1993, vandas2011, schmidt2012} or the termination shocks of a reconnection outflow jet \citep{aurass2002, mann2009}. However, an explanation of herringbone burstiness is still not complete. This bursting has been shown to have a quasi-periodicity of 2-11 seconds, at least for the 2011 September 22 event \citep{carley2013}. In this study we have shown that herringbones may have as much to do with the coronal environment as well as the shock acceleration mechanism, hence it may be the environment which produces this bursting. For example, a shock which quasi-periodically encounters changes in the coronal environment which promote the acceleration of particles could produce such a bursty radio signature. Alternatively, there are other mechanisms occurring in the corona which could lead to such bursty acceleration, such as a tearing mode instability in a current sheet \citep{kliem2000}. However, such a mechanism has only been attributed to particle acceleration observed at GHz frequencies, much higher frequency (much lower altitude) than the bursts we observe here. Further observation and modelling are needed to confirm or rule out any of the above mentioned mechanisms.

\section{Conclusion}

Herringbone radio bursts are direct signatures of particle acceleration occurring at coronal shocks. Given the high number of individual bursts, they provide a good opportunity to perform statistical analysis on the particle beam properties and/or the properties of the coronal environment into which the beams propagate. It is because of the high number of individual bursts that an automated routine to analyse them quickly and efficiently should be developed. 

We have shown here that the Hough transform can provide a good automated feature recognition routine to analyse herringbones, given that they appear as straight lines in the dynamic spectrum. However, the algorithm has some shortcomings, the most significant of which is failing to detect the start and stop frequencies of the bursts. As a results this step must be done by a human user, making the whole process semi--automated. Some extra processing steps are needed if all herringbone radio bursts are to be analysed for a complete study of their statistics in a fully automated way. {\color{black} This is especially pertinent given the latest generation of radio telescopes, such as the Low Frequency Array \citep[LOFAR;][]{vanHaarlem2013}. LOFAR's frequency range of ~20-240\,MHz, high sensitivity and high frequency and time-resolution are ideal for studying the detailed characteristics of a variety of fine-structure in solar radio bursts, such as herringbones, S-bursts and Type IIIs \citep{morosan2014, morosan2015}  -- we are set to make great gains in the knowledge of these phenomena with current and upcoming radio observational technology.}
 
We have shown here that a statistical analysis of herringbone fine structure can yield useful information on particle beam kinematics, the coronal environment that the bursts propagate in, and the emission mechanisms that are at play for these kinds of radio bursts. An analysis of a much larger set of herringbone bursts could yield important information in each of these areas, as well as produce a definitive theory of herringbone production, which does not yet exist.

\section*{Acknowledgments}
\noindent
Eoin Carley is supported by ELEVATE: Irish Research Council International Career Development Fellowship -- co-funded by Marie Curie Actions. 
Hamish Reid is supported by a SUPA Advanced Fellowship and an STFC grant ST/L000741/1. We would like to thank the referee for the 
useful and constructive comments.

\bibliographystyle{aa.bst}

\end{document}